\documentclass[%
 reprint,
 amsmath,amssymb,
 aps,
]{revtex4-2}

\newcommand{\be}{\begin{equation}}
\newcommand{\ee}{\end{equation}}
\newcommand{\bea}{\begin{eqnarray}}
\newcommand{\eea}{\end{eqnarray}}
\usepackage{graphicx}
\usepackage{dcolumn}
\usepackage{bm}
\usepackage{cancel}

\begin{document}

\preprint{APS/123-QED}

\title{Adiabatic invariant approach on Friedmann cyclic universe}

\author{Narakorn Kaewkhao}
 \altaffiliation{naragorn.k@psu.ac.th}
 \address{Department of Physics, Faculty of Science, Prince of Songkla University, Hatyai 90112, Thailand} 
 
\author{Phongpichit Channuie}
 \altaffiliation{channuie@gmail.com}
 \address{School of Science, Walailak University, Nakhon Si Thammarat, 80160, Thailand} 
\address{College of Graduate Studies, Walailak University, Nakhon Si Thammarat, 80160, Thailand}



\date{\today}

\begin{abstract}
Oscillating or cyclic models of the universe were inspired by Friedmann's seminal paper of 1922. The model supposes a closed universe. In this work, we study Friedmann closed universe using the adiabatic invariant approach. We start revisiting the cosmological force proposed by N. Rosen and derive the Lagrangian density from Rosen's concepts of cosmological force. Importantly, we introduce the Noether gauge symmetry followed by the Rund-Traumann identity (RTI) and adiabatic invariant approach to examine the cyclic models. We consider a single component form of relativistic matter (stiff matter) at $a(t=0)$ and $a(t=T)$, and surprisingly discover the world period is given by $\sim 15.8$ Gy which is very close to that computed using the Friedmann's formula for the cyclic universe. 
\end{abstract}


\maketitle


{\bf Introduction}: The general idea of a cyclic or oscillating universe went back to 1922. In Friedmann's seminal work, this was found that it was only formulated in a mathematically precise way \cite{Alexander1922}. The conclusion was that the physical demand of homogeneity and isotropy of space does not necessitate a static Universe. Friedman discovered a new class of non-static solutions of GR that is a generalization of Einstein’s $3$-dimensional hypersphere with a constant in space but changing in time curvature radius $R(t)$. Friedmann come up with the fundamental equation that governs the dynamics of the Universe. In his model, Friedmann assumed that life of the Universe is finite and his called this scenario the “periodic world”, see Ref.\cite{Belensiy2013} for more detailed discussion.

The adiabatic invariant approach has played an important role in atomic physics. It has been so far applied to the elliptical electron orbit case of the Sommerfeld model \cite{Sommerfeld1916}. In this work, we are interested in investigating the relation between the cosmological force first derived by Rosen, the adiabatic invariant proposed by Ehrenfest, and the closed universe by Friedmann. We starting reviewing the Rosen's idea about the cosmological force. We introduce the Noether gauge symmetry to express the Lagrangian density in the point-like form. We then consider the Rund-Traumann identity (RTI) and adiabatic invariant approach to examine the cyclic models. We focus on a single component of the universe in a form of relativistic matter (stiff matter) at the beginning and ending of the universe. We compute the world period of the Friedmann's cyclic universe.

{\bf Cosmological force revisit}: In this part, we begin with the proposal of Nathan Rosen regarding cosmological force.  Consider the fluid equation given by 
\be\label{fluid eq}
\dot{\rho}+3\frac{\dot{a}}{a}(\rho+\frac{p}{c^{2}})=0,
\ee
and the equation of state $p=w\rho c^{2}$. This gives the relation between the energy density and the scale factor, i.e., $\rho(a)=\rho_{0}a^{-3(1+w)}$. From the Friedmann equation (see Ref.\cite{BabaraRyden2017}, ch.4)
\be\label{Friedmann eq}
\frac{\dot{a}^{2}}{a^{2}}=\frac{8\pi G_{\rm N}}{3}\rho(t)-\frac{\kappa c^{2}}{R^{2}_{0}a^{2}}+\frac{\Lambda c^{2} }{3},
\ee
where three values of the curvature $\kappa=0, +1$ and $-1$ are set for spatially flat, open and closed universe respectively. The component of the universe due to the cosmological constant is $\rho_{\Lambda}=\frac{\Lambda c^{2}}{8\pi G_{\rm N}}.$ It should be noted that in some standard text on cosmology $k\equiv \frac{\kappa }{R^{2}_{0}}$(see Ref.\cite{WRinder}, ch.9) where $R_{0}=1.4\times 10^{26}\,m$ denotes for the present radius of the universe. Defining a critical density $\rho_{c}\equiv \frac{3c^{2}}{8\pi G_{\rm N}}H(t)^{2}$, we define the dimensionless density parameter as
\begin{eqnarray}
\Omega(t)\equiv \frac{\rho(t)}{\rho_{c}}\,.
\end{eqnarray}
Hence the Friedmann equation can be written as \cite{BabaraRyden2017}
\be\label{FM form}
\frac{\kappa c^{2}}{R^{2}_{0}}={a(t)^{2}H(t)^{2}}{\Big(\Omega(t)-1\Big)}.
\ee
Multiplying both sides of Eq.(\ref{Friedmann eq}) by $\frac{1}{2}ma^{2}$ yields
\bea\label{multiply FMeq}
\frac{1}{2}ma^{2}\frac{1}{a^{2}}\dot{a}^{2}=\frac{8\pi G_{\rm N}}{3}\rho(t)\frac{1}{2}ma^{2}&-&\frac{1}{2}\frac{m\kappa c^{2}}{R^{2}_{0}}-\frac{1}{6}ma^{2}\Lambda c^{2}\,,
\eea
or
\bea\label{multiplyFMeqa}
\frac{1}{2}m\dot{a}^{2}-\frac{4\pi G_{\rm N}ma^{2}\rho}{3}+\frac{1}{6}ma^{2}\Lambda c^{2}&=&-\frac{1}{2}\frac{\kappa mc^{2}}{R^{2}_{0}}\,.
\eea
Notice that the LHS of the above relation is $T+V$; while the RHS is $E$ where $T,\,V$ and $E$ represent the kinetic energy, the potential energy and the total energy of the system, respectively. It is worth to see the association the concept of force with the potential energy,
where $m$  can be seen as the test mass (passive mass) which is placed on the equipotential surface of self-gravitational system. 
\bea\label{potential term}
V(a)=-\frac{4\pi G_{\rm N}ma^{2}\rho(a)}{3}+\frac{1}{6}ma^{2}\Lambda c^{2}.
\eea
Therefore, we obtain the cosmological force governed by
\bea
F(a)&=&-\frac{\partial V(a)}{\partial a}=\frac{4\pi G_{\rm N}m}{3}\frac{d\Big( a^{2}\rho\Big)}{d a}-\frac{1}{3}ma\Lambda c^{2},
\eea
or
\bea
F(a)=\frac{4\pi G_{\rm N}m}{3}\Big( 2a\rho +a^{2}\frac{d\rho}{da} \Big)-\frac{1}{3}ma\Lambda c^{2}.\label{cosmological force}
\eea
Such a form of force was beautifully appeared in the work of Rosen \cite{Rosen1992}. Using the notion for  $\dot{\rho}=\frac{\partial \rho}{\partial a}\frac{\partial a}{\partial t}=-3\frac{\dot{a}}{a}(\rho+\frac{p}{c^{2}})$, we can write
$\frac{\partial \rho}{\partial a}=-\frac{3}{a}(\rho+\frac{p}{c^{2}})$. After substituting $\frac{\partial \rho}{\partial a}$ into Eq.(\ref{cosmological force}), this yields
\bea\label{force F}
F(a)&=&\frac{4\pi G_{\rm N}m}{3}\Big[ 2a\rho -\frac{3}{a}(\rho+\frac{p}{c^{2}})a^{2}-\frac{a\Lambda c^{2}}{4\pi G_{\rm N}} \Big]\nonumber\\
&=&\frac{4\pi G_{\rm N}m}{3}\Big[ 2a\rho -3a(\rho+\frac{p}{c^{2}})-\frac{a\Lambda c^{2}}{4\pi G_{\rm N}} \Big]\nonumber\\
&=&-\frac{4\pi G_{\rm N}ma}{3}\Big[ \rho+3\frac{p}{c^{2}}-\frac{\Lambda c^{2}}{4\pi G_{\rm N}} \Big]\nonumber\\
&=&m\ddot{a}.
\eea
We once reconfirm the acceleration equation or the Raychaudhuri equation  as follows:
 \begin{eqnarray}\label{acceleration eq}
 \frac{\ddot{a}}{a}=-\frac{4\pi G_{\rm N}}{3}(\rho+\frac{3p}{c^{2}})+\frac{\Lambda c^{2}}{3}.
 \end{eqnarray}
 It should be noted here as well that the acceleration equation can be obtained from the Euler-Lagrange equation for the  Lagrangian inspired from  Rosen's concept of cosmological force, i.e., 
\begin{eqnarray}
\mathcal{L}(a,\dot{a})&=&T-V\nonumber\\&=&\frac{1}{2}m\dot{a}^{2}+\frac{4\pi G_{\rm N}}{3}ma^{2}\rho(a)-\frac{1}{6}ma^{2}\Lambda c^{2}\,.\label{Nathan Lagrangian}
\end{eqnarray}

{\bf The Rund-Traumann identity and adiabatic invariance}: We discuss the Noether gauge symmetry approach for the system and quantify exact cosmological solutions with the help of
the Noether symmetries of point-like Lagrangian. We consider the first prolongation of NGS as
\begin{eqnarray}
X^{[1]}_{NGS}&=&X_{NGS}+\dot{\alpha}\frac{\partial}{\partial \dot{a}},\nonumber\\
&=& \Big( \tau\frac{\partial}{\partial t}+\alpha \frac{\partial}{\partial a}\Big)+\dot{\alpha}\frac{\partial}{\partial \dot{a}}
\end{eqnarray}
 where 
 \begin{eqnarray}
 \dot{\alpha}(t,a) &=& D_{t}\alpha-\dot{a}D_{t}\tau,\nonumber\\
 &=&\frac{\partial \alpha}{\partial t}+\dot{a}\frac{\partial \alpha}{\partial a}-\dot{a}\Big(\frac{\partial\tau}{\partial t}+\dot{a}\frac{\partial \tau}{\partial \dot{a}}\Big).
 \end{eqnarray}
 Using the Noether gauge symmetry condition given by
 \begin{eqnarray}
 X^{[1]}_{NGS}{\cal L}+{\cal L}D_{t}\tau &=&D_{t}B,
 \end{eqnarray}
 we have
 \begin{eqnarray}
 \tau\cancel{\frac{\partial L}{\partial t}}&+&\alpha\frac{\partial {\cal L}}{\partial a}
 +\Bigg(\frac{\partial \alpha}{\partial t}+\dot{a}\frac{\partial \alpha}{\partial a}-\dot{a}\Big(\frac{\partial\tau}{\partial t}+\dot{a}\frac{\partial \tau}{\partial \dot{a}}\Big)   \Bigg)\frac{\partial \mathcal{ L}}{\partial{\dot{a}}}\nonumber\\&+&\mathcal{L}\Big(\frac{\partial \tau}{\partial t}+\dot{a}\frac{\partial \tau}{\partial a}\Big)=\Big(\frac{\partial B}{\partial t}+\dot{a}\frac{\partial B}{\partial a}  \Big)\,.
 \end{eqnarray}
Note that the first term on the left-hand side is zero because Lagrangian as shown in Eq.(\ref{Nathan Lagrangian}) is not expressed in time variable explicitly. Hence we find
  \begin{widetext}
 \begin{eqnarray}
 \alpha\Big[\frac{8\pi G_{\rm N}ma\rho(a)}{3} +\frac{4\pi G_{\rm N}ma^{2}}{3}\frac{d\rho(a)}{da}&-&\frac{ma\Lambda c^{2}}{3}\Big]+\alpha_{t}(m\dot{a})+\alpha_{a}m\dot{a}^{2}-\tau_{t}m\dot{a}^{2}-m\tau_{a}\dot{a}^{3}+\frac{1}{2}m\dot{a}^{2}\tau_{t}+
 \frac{4\pi G_{\rm N}ma^{2}\rho(a)}{3}\tau_{t}\nonumber\\&-&\frac{ma^{2}\Lambda c^{2}}{6}\tau_{t}+\dot{a}\tau_{a}\frac{1}{2}m\dot{a}^{2}+
 \dot{a}\tau_{a}\Big(\frac{4\pi G_{\rm N}ma^{2}\rho(a)}{3}-\frac{ma^{2}\Lambda c^{2}}{6}\Big)=B_{t}+\dot{a}B_{a}.
 \end{eqnarray}
 \end{widetext}
 After separating monomial and polynomial, this gives the set of PDEs as
 \begin{widetext}
 \begin{eqnarray}
 \alpha\Big[\frac{8\pi G_{\rm N}ma\rho(a)}{3} +\frac{4\pi G_{\rm N}ma^{2}}{3}\frac{d\rho(a)}{da}-\frac{ma\Lambda c^{2}}{3}\Big]+\frac{4\pi G_{\rm N }ma^{2}\rho(a)}{3}\tau_{t}-\frac{ma^{2}\Lambda c^{2}}{6}\tau_{t}&=&B_{t},\\
 \alpha_{t}m+\tau_{a}\frac{4\pi G_{\rm N}ma^{2}\rho(a)}{3}-\tau_{a}\frac{ma^{2}\Lambda c^{2}}{6}&=&B_{a},\\
 \alpha_{a}m-\tau_{t}m+\tau_{t}\frac{m}{2}&=&0,\\
 -\frac{1}{2}m\tau_{a}&=&0\,.
 \end{eqnarray}
  \end{widetext}
It is rather straightforward to show that 
  \begin{widetext}
 \begin{eqnarray}
 \tau_{a}&=&0, \label{tau1}\\ 
 \alpha_{a}&=&\frac{\tau_{t}}{2},\\
 \alpha_{t}m&=&B_{a},\\
 \alpha\Big[\frac{8\pi G_{\rm N}ma\rho(a)}{3}-4\pi G_{\rm N}ma\rho(a)(1+w)  \Big]+\frac{4\pi G_{\rm N }ma^{2}\rho(a)}{3}\tau_{t}-\frac{ma^{2}\Lambda c^{2}}{6}\tau_{t}&=&B_{t}.\label{B1}
 \end{eqnarray}
\end{widetext}
 We once obtain the possible solution of Eq.(\ref{tau1})-Eq.(\ref{B1}) as shown below:
 \begin{eqnarray}
 \tau(t)&=&2c_{1}t+\tau_{0},\\
\alpha(a,t)&=&c_{1}a+c_{2}t+\alpha_{0},\\
B(a,t)&=&m(c_{2}a+c_{3}t)+B_{0},
 \end{eqnarray}
 where $\tau_{0},\alpha_{0}$ and $B_{0}$ are arbitrary constants.
 To simplify calculations , let's set $c_{2}=0,\alpha_{0}=0,$ and $\tau_{0}=0.$ Therefore, the new Lagrangian is expressed in term of  the linear transformation of $t$ and $a(t)$, i.e., $t'=t(1+2\epsilon c_{1})$ and $a'=a(1+\epsilon c_{1})$ which is inspired by Ref.\cite{Neuenschwander} given below 
 \begin{eqnarray}
 L'(a'(t'))=\frac{1}{2}m\dot{a'}^{2}+\frac{4\pi G_{\rm N}m{a'}^{2}\rho(a')}{3}-\frac{ma'^{2}\Lambda c^{2}}{6},
 \end{eqnarray}
 The Rund-Trautman identity (RTI)\cite{Neuenschwander}
 \begin{eqnarray}
\dot{\tau}&&\Big[ \mathcal{L}-\dot{q}^{i}(\partial \mathcal{L}/\partial \dot{q}^{i}) \Big] +\dot{\xi}^{i}(\partial \mathcal{L}/\partial \dot{q}^{i})+\tau(\partial \mathcal{L}/\partial t)\nonumber \\
 &&+\xi^{i}(\partial \mathcal{L}/\partial q^{i})=0.
 \end{eqnarray}
Actually, the Noether's theorem can be directly derived from the RTI by using the definition of the total derivative, i.e.
\begin{eqnarray}
\frac{d \mathcal{L}(t,q^{i},\dot{q}^{i})}{dt}=\frac{\partial \mathcal{L}}{\partial t}+
\frac{\partial \mathcal{L}}{\partial q^{i}}\frac{dq^{i}}{dt}+\frac{\partial \mathcal{L}}{\partial \dot{q}^{i}}\frac{d\dot{q}^{i}}{dt},
\end{eqnarray}
and the production rule for derivative, i.e. 
\begin{eqnarray}
p_{i}\frac{d\xi^{i}}{dt} &=&\frac{d(p_{i}\xi^{i})}{dt}-\xi^{i}\frac{dp_{i}}{dt}, \nonumber\\
&=&\frac{d(p_{i}\xi^{i})}{dt}-\xi^{i}\frac{d}{dt}\Big(  \frac{\partial \mathcal{L}}{\partial \dot{q}^{i}}  \Big),
\end{eqnarray}
where $p_{i}=\frac{\partial L}{\partial \dot{q}^{i}}.$
Taking Ref.\cite{Neuencschwander1993}, we can write Noether's theorem as follows: 
\begin{eqnarray}
&&(\dot{q}^{i}\tau-\xi^{i})\Big[ \partial \mathcal{L}/\partial q^{i}-d/dt(\partial \mathcal{L}/\partial \dot{q}^{i}) \Big]\nonumber \\&&=\frac{d}{dt}\Bigg\{\Big[\mathcal{L}-\dot{q}^{i}(\partial \mathcal{L}/\partial \dot{q}^{i})\Big]\tau+(\partial \mathcal{L}/\partial \dot{q}^{i})\xi^{i}  \Bigg\}.
\end{eqnarray}
If the Euler-Lagrange equation 
\begin{eqnarray}
\partial \mathcal{L}/\partial q^{i}-d/dt(\partial \mathcal{L}/\partial \dot{q}^{i})=0
\end{eqnarray}
is satisfied, then the Noether's corollary written in term of the canonical momenta $p_{i}$ and the Hamiltonian $\mathcal{H}$ as
 \begin{eqnarray}
 -\mathcal{H}\tau+p_{i}{\xi^{i}}= constant. \label{NTcore}
 \end{eqnarray}
The Hamiltonian $\mathcal{H}=q^{i}p_{i}-\mathcal{L}$. In this work, we set  $p_{a}=\frac{\partial L}{\partial \dot{a}}=m\dot{a}$ and $\xi^{a}=\alpha=c_{1}a$. the Hamiltonian function is
\begin{eqnarray}
\mathcal{H}(a,p_{a})=\frac{p_{a}^{2}}{2m}-\frac{4\pi G_{\rm N}}{3}ma^{2}\rho(a)+\frac{ma^{2}\Lambda c^{2}}{6}.
\end{eqnarray}
The constant of motion can be obtained from 
\begin{eqnarray}
\Sigma_{0}=\alpha\frac{\partial \mathcal{L}}{\partial \dot{a}}=\alpha m\dot{a}=c_{1}m a\dot{a}.
\end{eqnarray}
After replacing
\begin{eqnarray}
\tau(t)&=&2c_{1}t, \nonumber\\
\alpha&=&c_{1}a, \nonumber
\end{eqnarray}
to Eq.(\ref{NTcore}), it is easy to see that
\begin{eqnarray}
-\mathcal{H}t+\frac{1}{2}p_{a}a=constant.\label{adia form}
\end{eqnarray}
By taking  the average operation on both sides of Eq.(\ref{adia form}), we get that
$-\langle \mathcal{H}t\rangle+\frac{1}{2}\langle p_{a}a\rangle = constant. $
Following \cite{Neuencschwander1993}, in the context of cosmology, it was found that 
\begin{eqnarray}
\langle p_{a} a\rangle &=&m\langle a\dot{a}\rangle,\nonumber\\
&=& m\frac{\int_{0}^{T}a\dot{a}dt}{\int_{0}^{T}dt},\nonumber\\
&=&\frac{m}{T}\int_{0}^{T}a\frac{da}{dt}dt=\frac{m}{T}\int_{a(t=o)}^{a(t=T)}ada=0,
\end{eqnarray}
where we have set $a(0)=a(T)$ for the cyclic universe. According to the adiabatic invariant principle \cite{Ehrenfest1917}, it seems reasonable to assume that $\rho(a)$ vary slowly with time over one cycle \cite{adiabaticGleiser}, i.e., $|\dot{\rho}/\rho| \ll 1/T$. Since $|\dot{a}/a| \lesssim 1/T$, we can hence replace $\mathcal{H}$ with the average value $\langle \mathcal{H}t\rangle$. That yields
\begin{eqnarray}
\langle \mathcal{H}t\rangle &=&\frac{1}{T}\int_{0}^{T}\mathcal{H}t dt,\nonumber\\
&=& \frac{\langle \mathcal{H}\rangle} {T}\int_{0}^{T} tdt=\frac{1}{2}\langle \mathcal{H}\rangle T.
\end{eqnarray}
The Boltzmann's mechanical theorem can be generalized to
\begin{eqnarray}
\langle \mathcal{H}\rangle T=constant.
\end{eqnarray}
From the Legendre transformation, we have
\begin{eqnarray}
\int_{t_{i}}^{t_{f}}\mathcal{L}dt=\int_{t_{i}}^{t_{f}} p_{i}\dot{q}^{i}dt-\int_{t_{i}}^{t_{f}} \mathcal{H}dt.
\end{eqnarray}
Integrating over one cycle, this gives
\begin{eqnarray}
{\int_{0}^{T}\mathcal{L}dt} &=& \oint p_{a} da-\int_{0}^{T}\mathcal{H}dt,\nonumber\\
&=& \oint p_{a} da-\langle \mathcal{H}\rangle T.
\end{eqnarray}
Note that we use the definition of the time average, i.e.,
\begin{eqnarray}
\langle \mathcal{H}\rangle =\frac{\int_{0}^{T} \mathcal{H}dt}{T}.
\end{eqnarray}
It is easy to see that $\int_{0}^{T} \mathcal{H}dt=\langle \mathcal{H}\rangle T $. From the fact that 
$\int_{0}^{T} \mathcal{L} dt=0$, this gives
\begin{eqnarray}
\oint p_{a} da = \langle \mathcal{H}\rangle T = -\frac{1}{2}\frac{\kappa mc^{2}}{R^{2}_{0}}T, \label{HT}
\end{eqnarray}
where we have used $\langle \mathcal{H}\rangle=E$ given by Eq.(\ref{multiplyFMeqa}).

{\bf Confluence of two frameworks}: We combine two frameworks of Friedmann cyclic universe and Ehrenfest's adiabatic invariance. Due to $F(a)=\frac{d p_{a}}{dt}=m\frac{d\dot{a}}{dt}$, this leads to $p_{a}=\int F(a)dt$ where $F(a)$ is the cosmological force first proposed by Rosen. Therefore we can write
\begin{eqnarray}
\oint p_{a} da &=&-\oint \int\frac{4\pi G_{\rm N}ma}{3}\Big[ \rho+3\frac{p}{c^{2}} \Big]dt da,\nonumber\\
&=&- \frac{4\pi G_{\rm N}m}{3}( 1+3w)\int \Big( \oint a\rho(a)da\Big) dt,\nonumber\\
&=&- \frac{4\pi G_{\rm N}m}{3}( 1+3w)\rho_{0}\oint \Big(\int a^{-2-3w}da\Big)dt,\nonumber\\
&=& \frac{4\pi G_{\rm N}m}{3}\frac{\cancel{(1+3w)}}{\cancel{(1+3w)}}\rho_{0}{\oint a(t)^{-(1+3w)}dt} \label{cycle momentum}
\end{eqnarray}
where $\rho(a)=\rho_{0}a^{-3(1+w)}$ [See  Ref\cite{BabaraRyden2017} topic 5.1].
At the starting and ending points of the universe, it is reasonable to set $w=+1$ and $a=(\frac{t}{t_{0}})^{\frac{2}{3+3w}}$ for the relativistic matter (stiff matter). Then Eq.(\ref{cycle momentum}) becomes
\begin{eqnarray}
\oint p_{a}da &=& \frac{4\pi G_{\rm N}m\rho_{0}}{3}(-3)t_{0}^{4/3}T^{-1/3}_{\pi}=\langle \mathcal{H}\rangle T_{\pi}, \nonumber\\&=&-\frac{1}{2}\frac{\kappa mc^{2}T_{\pi}}{R^{2}_{0}}\,,
\end{eqnarray}
where $T_{\pi}$ denotes the period for cyclic universe. 
This implies that
\begin{eqnarray}
T_{\pi}&=& \Bigg[ \frac{8\pi G_{\rm N}\rho_{0}t_{0}^{4/3}R^{2}_{0}}{c^{2}} \Bigg]^{3/4}=\Big[ 8\pi G_{N}\rho_{0}  \Big]^{4/3}\Big[ \frac{R_{0}}{c} \Big]^{5/2}\nonumber\\
&\approx& 5\times 10^{17} s \times \frac{1 yr}{3.156\times 10^{7}\,s}=1.58\times 10^{10}\,yrs\,.
\end{eqnarray}
where $t_{0}=\frac{R_{0}}{c}=H_{0}^{-1}.$  
Interestingly, this numerical value computed using adiabatic invariance approach is in excellent agreement with that calculated using the Friedmann formula. Let's take a short recap about his formula. From a mathematical perspective of Friedmann, the radius of curvature would become a periodic function given by\cite{Alexander1922,ORaaifeartaighforgottenFMmodel,Belensiy2013} 
\begin{eqnarray}
T_{\pi}&=&\frac{2}{c}\int_{0}^{x_{0}}\sqrt{\frac{x}{A-x+\frac{\Lambda x^{3}}{3c^{2}}}}\,\, dx  \approx \frac{\pi A}{c}=\frac{8\pi G_{\rm N}}{c^{2}}\cdot\frac{M}{6\pi c}\nonumber\\&=&\frac{8 G_{\rm N}M}{6c^{3}}\approx 4.94\times 10^{17} s\approx  1.57\times 10^{10} yrs, \label{collapU}
\end{eqnarray}
 where $M$ is the total mass of the universe $M=\frac{1.5\times 10^{53} kg}{1.989\times 10^{30} kg }\approx 7.9\times 10^{22} M_\odot $\cite{PDavies2006}. The cosmic mass that we have considered is similar to that given in the Hoyle-Carvalho formula $M\sim c^{3}/(GH)\sim 10^{53} kg$ \cite{Dimitar2014,Carvalho,Srivaram}, $A=\frac{4G_{\rm N}M}{3\pi c^{2}}$ and $\Lambda$ is the cosmological constant. 
 
{\bf Conclusion}: In the present work, we have considered a formal framework of the adiabatic invariant approach in cosmology and applied to the Friedmann closed universe. We first derived the Lagrangian density from Rosen's concepts of cosmological force and then introduced the Noether gauge symmetry to the system. We used the Rund-Traumann identity (RTI) and adiabatic invariant approach to examine the cyclic models. The RTI and the Noether 's theorem provides a direct bridge between invariances and conserved quantities. We considered a single component of the universe in a form of relativistic matter (stiff matter) and discovered the world period is given by $\sim 15.8$ Gy. We found that this value is very close to that computed using the Friedmann's formula for the cyclic universe. However, in this exploratory study, our approach is minimalistic rather than aiming at great generality. Therefore, it can be generalized by considering a scalar field component of the universe. This would affect not only the totally mass of the universe but also the periodic world.      


{\bf Acknowledgements}: P. Channuie acknowledged the Mid-Career Research Grant 2020 from National Research Council of Thailand (NRCT5-RSA63019-03).



\end{document}